\documentclass[aps,prd,twocolumn,showpacs,floats,floatfix,letterpaper,nofootinbib,superscriptaddress,preprintnumbers]{revtex4}

\usepackage{amssymb,amsmath,latexsym,mathrsfs}
\usepackage{graphicx}
\usepackage{epsfig}
\usepackage{multirow}
\usepackage{array}
\usepackage{xspace}
\usepackage{color}

\newcommand{\bea}{\begin{eqnarray}}
\newcommand{\eea}{\end{eqnarray}}

\newcommand{\ms}{m_{s}^{\rm{eff}}}
\newcommand{\neff}{N_{\textrm{eff}}}

\newcommand{\mnu}{{\Sigma}m_{\nu}}

\begin{document}


\title{Cosmological limits on neutrino unknowns versus low redshift priors}

\author{Eleonora Di Valentino}
\affiliation{Institut d'Astrophysique de Paris (UMR7095: CNRS \& UPMC- Sorbonne Universities), F-75014, Paris, France}
\author{Elena Giusarma}
\affiliation{Physics Department and INFN, Universit\`a di Roma ``La Sapienza'', Ple Aldo Moro 2, 00185, Rome, Italy}
\author{Olga Mena}
\affiliation{IFIC, Universidad de Valencia-CSIC, 46071, Valencia, Spain}
\author{Alessandro Melchiorri}
\affiliation{Physics Department and INFN, Universit\`a di Roma ``La Sapienza'', Ple Aldo Moro 2, 00185, Rome, Italy}
\author{Joseph Silk}
\affiliation{Institut d'Astrophysique de Paris (UMR7095: CNRS \& UPMC- Sorbonne Universities), F-75014, Paris, France}
\affiliation{AIM-Paris-Saclay, CEA/DSM/IRFU, CNRS, Univ. Paris VII, F-91191 Gif-sur-Yvette, France}
\affiliation{Department of Physics and Astronomy, The Johns Hopkins University Homewood Campus, Baltimore, MD 21218, USA}
\affiliation{BIPAC, Department of Physics, University of Oxford, Keble Road, Oxford
OX1 3RH, UK}
\begin{abstract}
Recent Cosmic Microwave Background (CMB) temperature and polarization anisotropy measurements from the Planck mission  have significantly improved previous  constraints on the neutrino masses as well as the bounds on extended models with massless or massive sterile neutrino states. However, due to parameter degeneracies, additional low redshift priors are mandatory in order to sharpen the CMB neutrino bounds. 
We explore here the role of different priors on low redshift quantities, such as the Hubble constant, the cluster mass bias, and the reionization optical depth $\tau$. Concerning current priors on the Hubble constant and the cluster mass bias, the bounds on the neutrino parameters may  differ appreciably depending on the choices adopted in the analyses. With regard to  future improvements in the priors on the reionization optical depth, a value of $\tau=0.05\pm 0.01$, motivated by astrophysical estimates of the reionization redshift, would lead to $\sum m_\nu<0.0926$~eV at $90\%$~CL, when combining the full \textit{Planck} measurements, Baryon Acoustic Oscillation and Planck clusters data, thereby opening  the window to unravel the neutrino mass hierarchy with existing cosmological probes.  
\end{abstract}
\preprint{IFIC/15-84}
\maketitle

\section{Introduction}
Recent Plank measurements of the Cosmic Microwave Background (CMB) temperature and polarization anisotropy offer a unique test of some particle properties  which still remain unknown~\cite{Ade:2015xua}. The absolute neutrino masses and their ordering are among the most frequently exploited topics in the literature (see e.g. Refs.~\cite{Ade:2015xua,DiValentino:2015wba,Giusarma:2014zza,Gerbino:2015ixa}). The most stringent bound quoted from the Planck collaboration, combining their CMB measurements with Baryon Acoustic Oscillations (BAO) is $\sum m_\nu < 0.17$~eV at $95\%$~CL~\cite{Ade:2015xua}. However, this is certainly not the most restrictive constraint to date. Measurements of Lyman $\alpha$ absorption in distant quasar spectra tighten the previous bound to $\sum m_\nu< 0.12$~eV at $95\%$~CL~\cite{Palanque-Delabrouille:2015pga}, based on new hydrodynamical simulations, especially devoted to keeping systematic uncertainties under control. Therefore, the role of low redshift observables (such as the Lyman $\alpha$ forest, the BAO signal, the Hubble constant $H_0$ or the constraints from the cluster redshift distribution) is crucial, as these measurements help enormously in pinning down the CMB neutrino mass constraint. Among these possible external data sets, we focus here on direct measurements of the Hubble constant $H_0$ and the cluster number counts, as their constraining power on $\sum m_\nu$ strongly depends of the choice of priors. Concerning the former, there are currently at least two possible $H_0$ measurements one may apply, which would lead to different $\sum m_\nu$ constraints. Regarding the latter, the prior on the cluster mass bias is a critical quantity which could even lead to non-zero neutrino masses. We will explore here the different neutrino mass bounds that are obtained with the possible prior choices on $H_0$ and on $1-b$, the cluster mass bias. Furthermore, we shall also illustrate the impact on neutrino mass bounds from a near future and improved prior on an additional low redshift quantity, the reionization optical depth $\tau$. The prior used here focus on a lower value of $\tau$, and it is motivated by hints from high-redshift quasar absorption and Lyman $\alpha$ emitters.

The structure of the paper is as follows. We start in Sec.~\ref{sec:method} with a description of the three different cosmological models explored here, which account for different neutrino parameters. The basic cosmological data sets used in our analyses are also detailed in this first section. Section~\ref{sec:anal} presents the results of our numerical analyses in each of the three neutrino scenarios considered here, focusing on the role of the Hubble constant, cluster mass bias, and reionization optical depth priors. We conclude in Sec.~\ref{sec:concl}. 

\section{Methodology}
\label{sec:method}
We analyze here three different scenarios, by varying the following set of parameters:
\begin{equation}\label{parameter}
\{\Omega_{\textrm{b}}h^2,\Omega_{\textrm{c}}h^2, \Theta_s, \tau, n_s, \log[10^{10}A_{s}], \sum m_\nu, \ms, \neff,\}~,
\end{equation}
where we have the six parameters of the $\Lambda$CDM model, i.e.  the baryon $\Omega_{\textrm{b}}h^2$ and the cold dark matter $\Omega_{\textrm{c}}h^2$ energy densities, the ratio between the sound horizon and the angular diameter distance at decoupling $\Theta_{s}$, the reionization optical depth $\tau$, and the inflationary parameters, the scalar spectral index $n_s$ and the amplitude of the primordial spectrum $A_{s}$. Moreover, we allow for  variations in this model, exploring three different scenarios, enlarging by one extra parameter each model. We first consider a $\Lambda$CDM model plus neutrino masses ($\sum m_\nu$), then we also consider the possibility of having additional relativistic degrees of freedom ($\sum m_\nu$ and $N_{\rm eff}$, with $N_{\rm eff}$-3.046 extra relativistic species), and lastly, we consider the possibility of massive sterile neutrinos ($\sum m_\nu$, $N_{\rm eff}$ and $\ms$, with $N_{\rm eff}$-3.046 extra massive species with a mass $\ms$). We have assumed that active neutrinos have a degenerate mass spectrum, with a minimum value $\sum m_\nu=0.06$~eV, as indicated by neutrino oscillation data. In principle, one could also consider the lightest neutrino mass eigenstate as the free parameter (instead of $\sum m_\nu$), and derive, making use of the neutrino mass splittings, the total neutrino mass. In such a case, two different runs, one for the normal hierarchy, and a separate one for the inverted hierarchy, would be needed. However, the bounds presented here will not change much in this situation, given the data sets exploited, which are mostly only sensitive to the total neutrino mass, and not to the hierarchical structure of the neutrino mass, i.e. their mass distribution (see, for example, \cite{Slosar:2006xb}).

For all these parameters, we use the flat priors listed in Table~\ref{tab:priors}. 

\begin{table}
\begin{center}
\begin{tabular}{c|c}
Parameter                    & Prior\\
\hline
$\Omega_{\textrm{b}}h^2$         & $[0.005,0.1]$\\
$\Omega_{\textrm{c}}h^2$       & $[0.001,0.99]$\\
$\Theta_{\rm s}$             & $[0.5,10]$\\
$\tau$                       & $[0.01,0.8]$\\
$n_s$                        & $[0.8, 1.2]$\\
$\log[10^{10}A_{s}]$         & $[2,4]$\\
$\sum m_\nu$ (eV)               & $[0.06,3]$\\
$\ms$  (eV) & [0,3]\\
$N_{\rm eff}$ & [3.046,10]\\
\end{tabular}
\end{center}
\caption{External priors on the cosmological parameters assumed in this paper.}
\label{tab:priors}
\end{table}

\subsection{Cosmological data}
We constrain the cosmological parameters previously described by using several combination of data sets. Our CMB measurements are those from the full Planck 2015 release on temperature and polarization  CMB angular power spectra~\cite{Adam:2015rua,Ade:2015xua}. The large angular scale temperature and polarization measured by the Planck LFI experiment is combined with the small-scale \emph{TT} temperature spectrum measured by Planck HFI, and we refer to this data set as \textit{Planck}. Moreover, when adding to this combination the small-scale \emph{TE} and \emph{EE} polarization spectra measured by Planck HFI, we shall refer to this data set as \textit{Planck pol}.

We consider also measurements of the large scale structure of the universe in their geometrical form, the Baryon Acoustic Oscillations (BAO) data. We include the 6dFGS \cite{Beutler:2011hx}, SDSS-MGS \cite{Ross:2014qpa},  BOSS LOWZ \cite{Anderson:2013zyy} and CMASS-DR11 \cite{Anderson:2013zyy} measurements as in \cite{Ade:2015xua}, referring to the combination of all of them as \textit{BAO}.

Then, we study the impact of the most relevant low redshift priors (concerning neutrino physics limits). First, we impose five different gaussian priors on the Hubble constant. Then, we consider the second Planck cluster catalog obtained through the Sunyaev-Zel'dovich (SZ) effect, analysing the impact of the different cluster mass biases, referring to this data set as \textit{SZ}. Finally, we study the effect of lowering the prior on the reionization optical depth $\tau$, as preferred by astrophysical measurements. In particular, we use two gaussian priors, $\tau=0.06\pm0.01$ and $\tau=0.05\pm0.01$.

Our constraints are obtained making use of the latest available version of the Monte Carlo Markov Chain (MCMC) package \texttt{cosmomc} \cite{Lewis:2002ah,Lewis:2013hha} with a convergence diagnostic based on the Gelman and Rubin statistics. This includes the support for the Planck data release 2015 Likelihood Code \cite{Aghanim:2015wva} implementing an optimal sampling~\cite{Lewis:2013hha}. The foreground parameters are varied as in Refs.~\cite{Ade:2015xua,Aghanim:2015wva}.

\section{Low-redshift priors}
\label{sec:anal}
\subsection{Hubble constant priors}
We consider here five possible constraints on the Hubble constant $H_0$,
without making any preference for one value over another.
The goal of our paper is indeed to discuss the impact of these different
priors on neutrino physics without entering the current debate
if one prior is more reliable than another. The first prior on $H_0$ arises from the recalibration of the authors of Ref.~\cite{Humphreys:2013eja} combined with the original Hubble Space Telescope (HST) measurements~\cite{Riess:2011yx}, which leads to the value of $H_0=73.0\pm 2.4$ km s$^{-1}$ Mpc$^{-1}$, hereafter $H073p0$ (see also Refs.~\cite{Bennett:2014tka,Cuesta:2014asa}). The second and the third possible choices exploited here for the prior on the Hubble constant arise from a recent reanalysis of ~\cite{Efstathiou:2013via}. One consists of a value $H_0=70.6\pm 3.3$ km s$^{-1}$ Mpc$^{-1}$ (hereafter $H070p6$), in better agreement with Planck 2015 findings, which has been dubbed as a \emph{conservative} estimate of the Hubble constant. The other value is $H_0=72.5\pm 2.5$ km s$^{-1}$ Mpc$^{-1}$ (hereafter $H072p5$). For the fourth and fifth $H_0$ constraints, we shall consider the values obtained by~\cite{Rigault:2014kaa}, correlating the host galaxy with the intrinsic luminosity. In particular, the priors are: $H_0=70.6\pm 2.6$ km s$^{-1}$ Mpc$^{-1}$ (hereafter $H070p6ref$), derived when using Cepheid distances
calibrated to the megamaser NGC 4258, Milky Way parallaxes, and LMC distance, and $H_0=68.8\pm 3.3$ km s$^{-1}$ Mpc$^{-1}$ (hereafter $H068p8$) obtained when using Cepheid distances calibrated solely with the distance to the NGC 4258 megamaser. In the following we shall explore the impact of all these possible priors on the neutrino parameters, without preferring one value over another, in order to avoid biases due to the choice. In fact, even if there is a tension at about 2$\sigma$ between some of them, in particular $H073p0$, and the $H_0$ value we have from Planck data, we have no clear justification at the moment to consider them affected by systematics. Furthermore, the tension is considerably reduced when varying the neutrino effective number $\neff$. Moreover, the constraints obtained combining Planck pol with $H073p0$ are consistent with those ones obtained with BAO measurements, where there is no indication of bias. We warn however again the reader to do not immediately consider the constraints that contains the $H073p0$ prior at the same level of fidelity of those based on the inclusion of the BAO dataset or of more conservative Hubble constant priors, especially when they are ruling out at $95\%$~CL significant regions of the neutrino parameter space. 

\begin{figure}[!t]
\includegraphics[width=8.5cm]{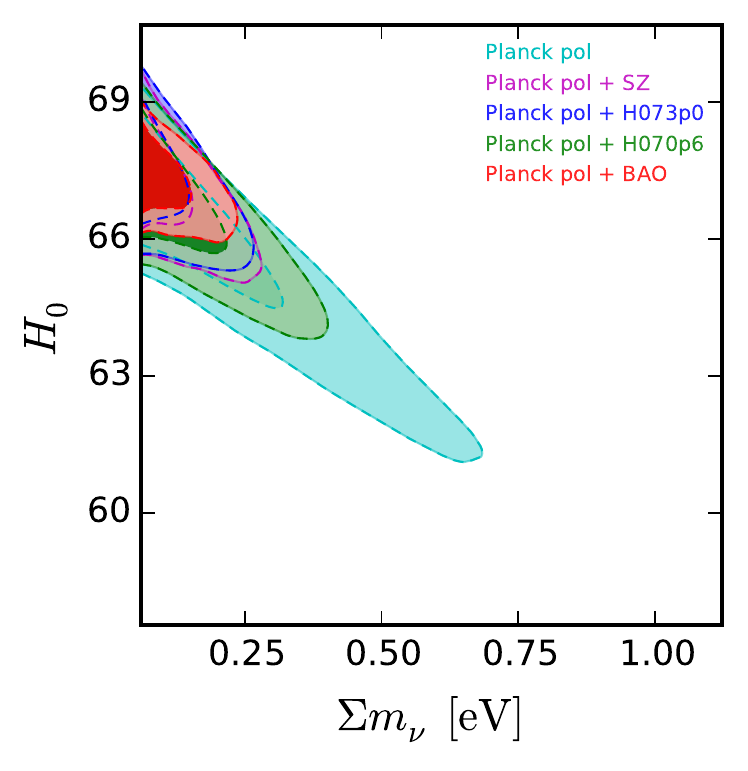} 
\caption{$68\%$ and $95\%$~CL allowed regions in the ($\sum m_\nu$, $H_0$) plane illustrating the effect of the low redshift priors studied here.}
\label{fig:fig1a}
\end{figure}

\begin{figure}[!t]
 \includegraphics[width=8.5cm]{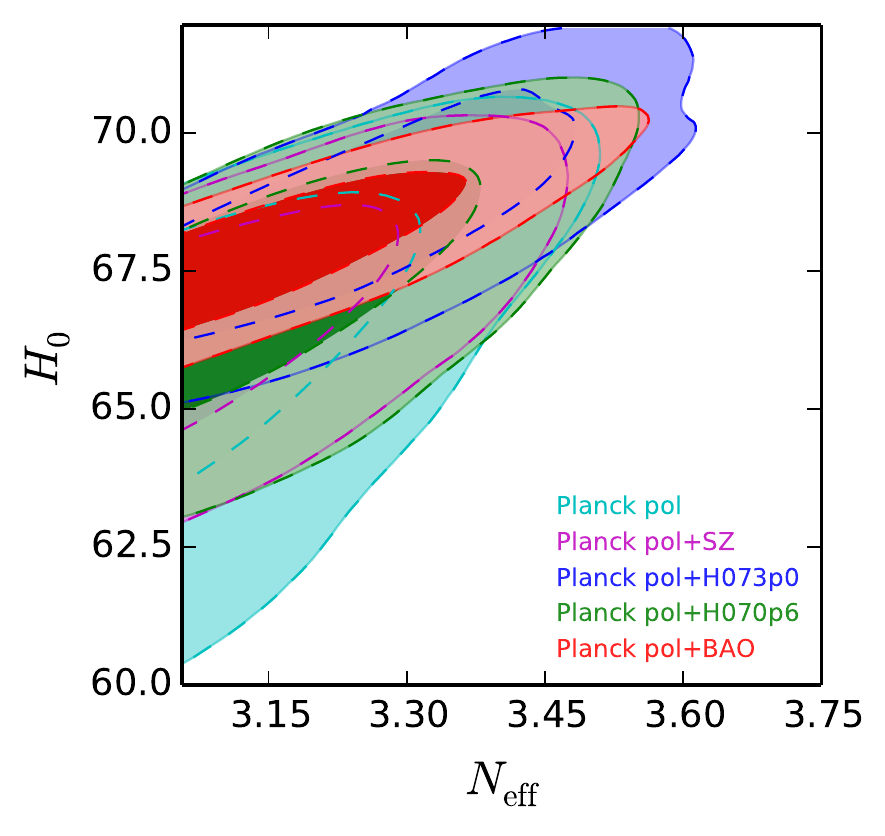}
\caption{As Fig.~\ref{fig:fig1a} panel but extending the neutrino mass model with $\neff$ dark radiation species, illustrating the ($\neff$, $H_0$) plane.}
\label{fig:fig1b}
\end{figure}

There exists a strong, well-known degeneracy between the neutrino mass and the Hubble constant (see e.g. ~\cite{Giusarma:2012ph} and Fig.~\ref{fig:fig1a}). In the absence of an independent measurement of $H_0$, the change in the CMB temperature anisotropies induced by the presence of massive neutrinos (which shifts the location of the angular distance to the last scattering surface) can be easily compensated by a smaller value of the Hubble constant. Therefore, the parameters $\sum m_\nu$ and $H_0$, when extracted from CMB data only, exhibit a very large degeneracy, see the cyan contours of Fig.~\ref{fig:fig1a}. Indeed, the $95\%$~CL bound on $\sum m_\nu$ from $Planck$ data set is $0.754$~eV. The addition of high multipole polarisation data (i.e. \textit{Planck pol}) leads to $\sum m_\nu< 0.497$~eV at $95\%$~CL. The reason for this improvement is due to the fact that polarization measurements alleviate many parameter degeneracies, among others, the $m_\nu$--$\tau$ degeneracy, with $\tau$ the reionization optical depth.  The associated mean value of $H_0=66.3$~km s$^{-1}$ Mpc$^{-1}$, see Table~\ref{tab:mah0}, is considerably smaller than the value quoted by the Planck collaboration within the $\Lambda$CDM model scenario with $\sum m_\nu=0.06$~eV ($H_0=67.3$ km s$^{-1}$ Mpc$^{-1}$) due to the degeneracy with the neutrino mass. Notice, from Tables~\ref{tab:mah0} and~\ref{tab:mah1}, that adding a prior on $H_0$ improves enormously the bounds on $\sum m_\nu$. The addition of the $H073p0$ prior has a much larger impact than the other ones, since it is associated to a larger Hubble constant, and this quantity is anti-correlated with $\sum m_\nu$. The $95\%$~CL on the total neutrino mass is $0.180$~eV. From the results from our MCMC analyses (some of them not depicted in Table~\ref{tab:mah0}) we conclude that the only data combination which provides competitive neutrino mass limits to those obtained  with CMB measurements plus the $H073p0$ prior is the one obtained combining Planck with BAO data. Both full-shape halo measurements and the others priors on the Hubble constant lead to weaker neutrino mass constraints.

Concerning the effective number of relativistic species $\neff$, the addition of Planck polarisation measurements leads to a major improvement in its $95\%$~CL bound, see Table~\ref{tab:maneffh0}, in which we show the results within the $\sum m_\nu$ + $\neff$ model for the same data combinations of Table~\ref{tab:mah0}. The constraints on the total neutrino mass are less restrictive in this more general scenario, given the strong degeneracy between $\sum m_\nu$ and $\neff$: a larger matter density can be compensated with an extra radiation component, and, consequently,  
$\sum m_\nu$ and $\neff$ are positively correlated. Figure~\ref{fig:fig1b} shows that the Hubble constant $H_0$ and $\neff$ are also positively correlated,  as the shift induced in the matter-radiation equality era by a larger $\neff>3.046$ can in principle  be compensated with a larger value of the Hubble constant, assuming that the matter-radiation equality redshift and the angular size of the horizon at recombination are free parameters. The measurements from Planck of the CMB damping tail alleviate this degeneracy (as  a value of $\neff>3.046$ will induce a higher expansion rate, which is translated into an increased Silk damping at high multipoles $\ell$). This is clear from the results shown in Table~\ref{tab:maneffh0}, where it can be noticed that the addition of the $H070p6$ prior to Planck measurements results in values of $\neff$ and $H_0$ which are considerably smaller than the ones obtained with the $H073p0$ prior. Interestingly, the values obtained in the $H070p6$ case are in very good agreement with those found when considering Planck plus BAO data, both in the cases of $Planck$ and \textit{Planck pol} data sets. The combination of $Planck$ data plus the $H073p0$ prior allows for the presence of an extra sterile neutrino at the $\sim 2\sigma$ level. Therefore, polarisation measurements play a major role in the constraints on the $\neff$ parameter, as once that they are considered, the bounds on $\neff$ become more robust and almost independent of the external priors. 

However, the extra neutrino species could also be massive, as motivated by the so-called neutrino oscillation anomalies~\cite{Abazajian:2012ys}. Massive sterile neutrinos do not necessarily need to have thermal abundances at decoupling, as their abundance is determined by their mixings with the active neutrino states~\cite{Melchiorri:2008gq}. In the following, we shall constrain simultaneously the $\neff$ massive sterile neutrino scenario and the sum of the three active neutrino masses $\sum m_\nu$. Therefore, the number of massive sterile neutrino species is given by $\Delta\neff=\neff- 3.046$, and their mass is $m^\textrm{eff}_s$. This mass is related to the physical mass by:
\begin{equation}
\label{parameter}
m^\textrm{eff}_s= (T_s/T_\nu)^3m_s=(\Delta \neff)^{3/4} m_s~,
\end{equation}
in which $T_s$ ($T_\nu$) is the current temperature of the sterile (active) neutrinos, and we have assumed that the sterile states have a phase-space distribution similar to that of the active neutrino states. Table~\ref{tab:maneffmsh0} shows the results for $\sum m_\nu$, $\neff$, $\ms$ and the other cosmological parameters previously considered as well. Notice that, in general, while the values of  $\neff$ and $\mnu$ are very similar to those obtained in the previous scenario, the value of the clustering parameter $\sigma_8$ is always reduced, as there is another source of suppression of the large scale structure growth, the sterile neutrino mass. 
 We will see in the next section that the inclusion of the clustering data, that mostly constrain the clustering parameter $\sigma_8$ and the current universe's matter density $\Omega_m$, can help to break these degeneracies.

Concerning the reionization optical depth, its value is always increased with respect  to its values in the other two previous neutrino mass models. The reason for that is due to the suppression of power on small scales induced by the presence of neutrino masses, an effect which can be compensated by increasing the amplitude of the primordial spectrum $A_s$. From CMB temperature data there exists a strong degeneracy between $A_s$ and $\tau$ (as long as the factor $A_s e^{-2\tau}$ is kept constant), which is broken, albeit only partially, by polarisation measurements. A higher value of $A_s$ can in turn be compensated by a larger $\tau$, and therefore the larger the total neutrino mass is (from both active and sterile states), the larger the reionization optical depth should be.  The tightest constraints in the sterile neutrino effective mass are obtained, as expected, after applying the $H073p0$ prior, since the Hubble constant is anti-correlated with both the active and the sterile neutrino masses. 

\begin{figure}[!t]
\includegraphics[width=8.5cm]{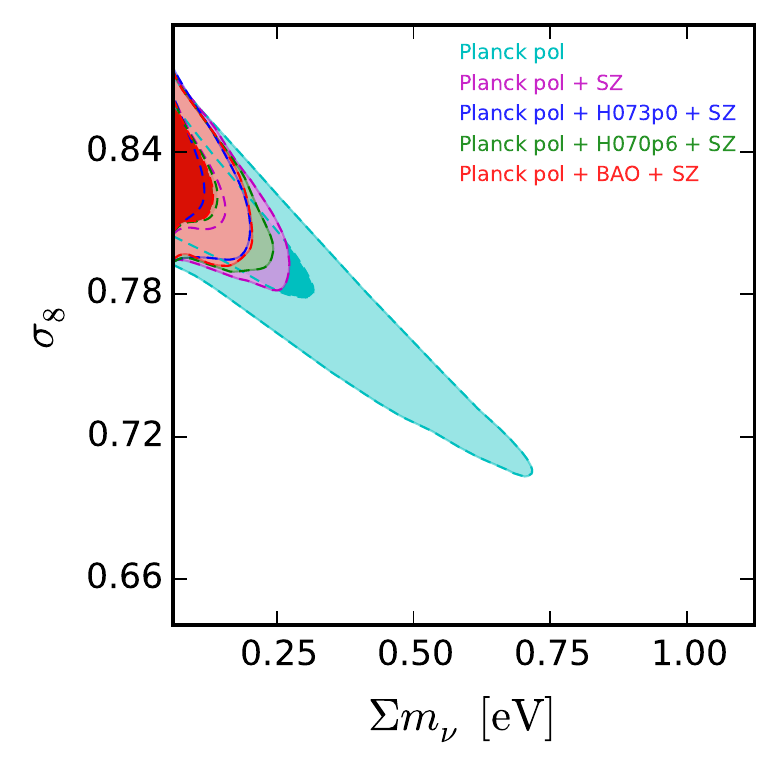} 
\caption{$68\%$ and $95\%$~CL allowed regions in the ($\sum m_\nu$, $\sigma_8$) plane illustrating the effect of the low redshift priors studied here.}
\label{fig:fig2a}
\end{figure}

\begin{figure}[!t]
\includegraphics[width=8.5cm]{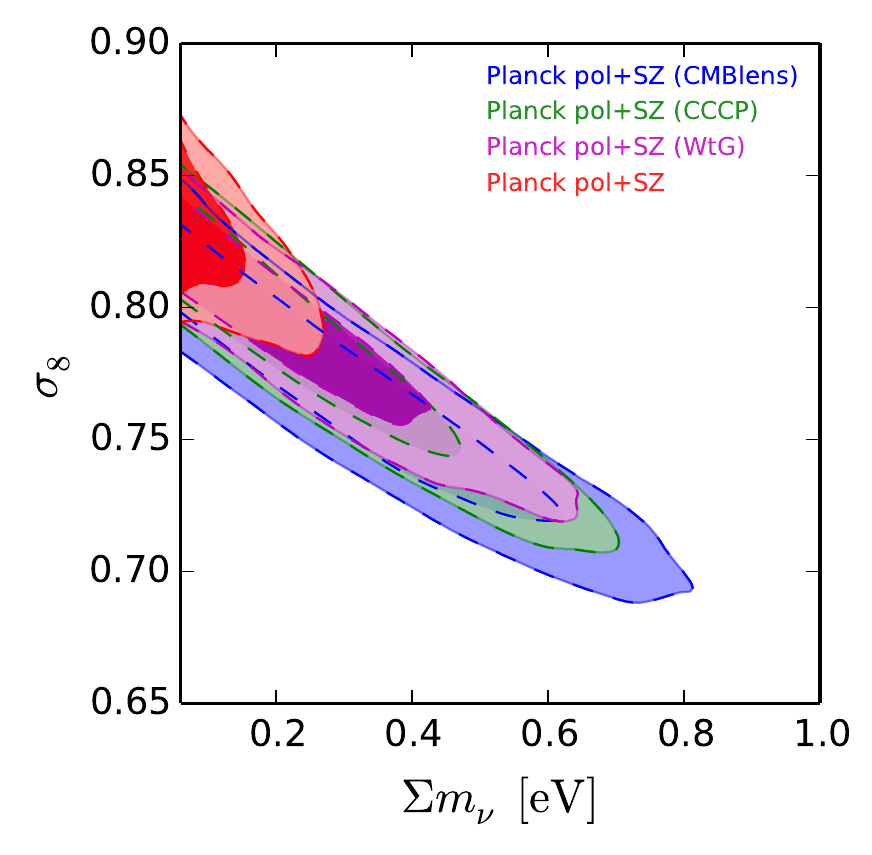} 
\caption{$68\%$ and $95\%$~CL allowed regions in the ($\sum m_\nu$, $\sigma_8$) plane, focusing on the impact of the cluster mass bias prior.}
\label{fig:figcluster}
\end{figure}

\begin{figure*}[!t]
\begin{center}
\includegraphics[width=1\textwidth]{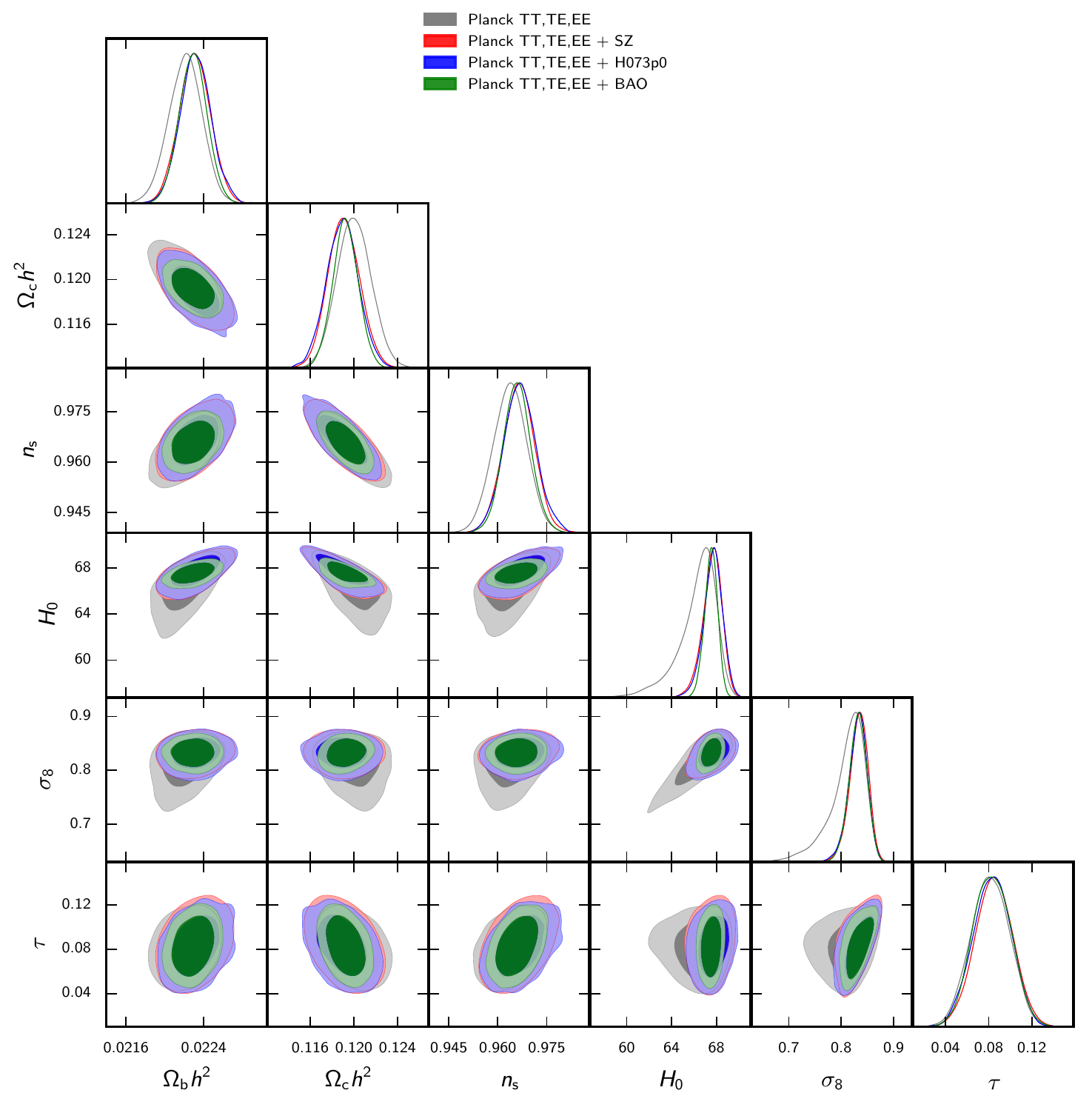} 
\caption{$68\%$ and $95\%$~CL allowed regions in the combined two-dimensional planes for the parameters $\Omega_{\textrm{c}}h^2$, $n_s$,$H_0$, $\sigma_8$ and $\tau$, together with their one-dimensional posterior probability distributions, arising from the combination of \textit{Planck pol} plus BAO, \textit{Planck pol} plus $H073p0$ and \textit{Planck pol}  plus SZ measurements.}
\label{fig:tri}
\end{center}
\end{figure*}

\subsection{Planck SZ Clusters}

The largest virialized objects in the universe are clusters of galaxies, providing  a unique way to extract the cosmological parameters. Cluster surveys usually focus on the cluster number count  function $d N/d z$, which measures the number of clusters of a certain mass $M$ over a range of redshift:
\bea
{d N\over dz}\Big|_{M>M_{\rm min}}=f_{\rm sky} {dV(z)\over dz}\int_{M_{\rm min}}^\infty dM \,{dn\over dM}(M, z)~,
\eea
with $f_{\rm sky}=\Delta\Omega/4\pi$ the fraction of sky covered by the survey and $\frac{dV(z)}{dz}$ the differential volume, which reads as
\bea
\frac{dV(z)}{dz}=\frac{4\pi}{H(z)} \int_0^z dz' \left(\frac{1}{H(z')}\right)^2~.
\eea

The cluster number count function is then related to its predictions within an underlying cosmological model. The main uncertainties arise from the cluster mass,  determined through four main available methods: X-rays, velocity dispersion, SZ effect, and weak lensing. Therefore, a crucial parameter in the analyses is the so-called cluster mass bias factor $1-b$, which accounts for deviations between the inferred X-ray cluster mass and the true cluster mass due to cluster physics and observational and/or selection effects. The overall error in the cluster mass determination is usually around  $\Delta M/M\sim 10\%$.  We exploit here the Planck Sunyaev-Zeldovich (SZ) 2015 cluster catalog, which consists of 439 clusters~\cite{Ade:2015gva,Ade:2015fva}. As we shall see in the following, the prior assumptions on the cluster mass bias $1-b$ (assumed to be a constant) play a major role in the neutrino mass constraints. Tables~\ref{tab:mapsz} and~\ref{tab:mapsz2} present the constraints on the neutrino mass and on a set of cosmological parameters, previously considered as well. The prior on the cluster mass bias quantity $1-b$ is freely varied in the $[0.1,1.3]$ range. Notice that, for this choice of the cluster mass bias prior, the $95\%$~CL neutrino mass limit after combining with $Planck$ is $0.206$~eV, which is further reduced down to  $0.184$~eV when polarisation measurements are also considered in the analysis. 
Furthermore, we find for these cases $1-b=0.656\pm0.051$ and $1-b=0.638\pm0.040$ at $68\%$~CL, respectively. 
Cluster number counts mostly constrain the clustering parameter $\sigma_8$ and the current universe's matter density $\Omega_m$, both involved in the calculation of the cluster mass function $d n(z, M)/d M$ through $N$-body simulations~\cite{Tinker:2008ff}. For the first case considered here, in which the cluster mass bias is a free parameter, the mean values of the parameter $\sigma_8$ obtained in the massive neutrino scenario are close to those obtained in the simple $\Lambda$CDM scenario with $\sum m_\nu =0.06$~eV, and therefore one can expect very tight neutrino mass bounds. Figure \ref{fig:fig2a} illustrates the strong degeneracy between the neutrino mass $\sum m_\nu$ and the clustering parameter $\sigma_8$ for several of the data combinations considered here. The tightest $95\%$~CL neutrino mass constraint we find is $\sum m_\nu<0.126$~eV, arising from the combination of \textit{Planck pol}, BAO, $H073p0$ and SZ data. As these three data sets (BAO, $H073p0$ and SZ) show no tension in the extraction of the different cosmological parameters (see Fig. \ref{fig:tri}), the neutrino mass bound arising from their combination and above quoted should be regarded as a robust limit.

Nevertheless, there exist lensing estimates of the $1-b$ parameter, which we shall also exploit in the following~\cite{Ade:2015fva}. The first two cluster mass bias priors arise from gravitational shear measurements from the Weighing the Giants (WtG)~\cite{vonderLinden:2014haa} and the Canadian Cluster Comparison Project (CCCP)~\cite{Ade:2015fva}, which lead to $1-b=0.688\pm 0.072$ and $0.780\pm 0.092$, respectively. CMB lensing offers yet another way of estimating the cluster masses~\cite{Zaldarriaga:1998te}, leading to a constraint on $1/(1-b)=0.99\pm 0.19$~\cite{Melin:2014uaa}. Table~\ref{tab:macluster} shows the $95\%$~CL limits on the neutrino mass as well as the mean values and $95\%$~CL associated errors on the remaining cosmological parameters explored here for each of the three possible cluster mass bias from lensing considerations. Notice that the values of the clustering parameter $\sigma_8$ are smaller, lying $1-2\sigma$ away  from the values obtained in the case of a freely varying cluster mass bias $1-b$. A lower value of $\sigma_8$ implies smaller clustering. A larger value of the neutrino mass would therefore be favoured, in order to suppress the small-scale clumping. Indeed, from the results depicted in Table~\ref{tab:macluster}, one can notice that the CMB lensing cluster mass prior is the one which leads to the largest bounds on the value of the total neutrino mass, as it suggests the lowest value of $\sigma_8$. Nevertheless, the CMB and SZ data combination do not find evidence for a non-zero neutrino mass. The impact of the cluster mass bias prior in the neutrino mass constraints can be also inferred from the results depicted in Fig.~\ref{fig:figcluster}, where it can be noticed that the four weak lensing cluster mass bias priors lead to a much larger degeneracy than for the free prior case, indicating a tension between primary CMB and SZ measurements of $\sigma_8$. The most extreme case corresponds to the CMBlens case, in which clusters would be much less massive than what primary CMB data seems to indicate, pointing therefore to a small value of $\sigma_8$, which in turn is translated into a relatively loose $95\%$~CL constraint of $\sum m_\nu< 0.669$~eV. On the other hand, the WtG cluster mass bias prior would indicate more massive clusters, and therefore the former limit is slightly tightened to  $\sum m_\nu< 0.531$~eV, at $95\%$~CL. 

The effects of the SZ prior in extended models (in which additional massless or massive species  are also considered) are shown in Tables~\ref{tab:maneffpsz} and \ref{tab:maneffmspsz}, from which we notice that, as in the $\Lambda$CDM-$\sum m_\nu$ scenario, the most constraining data set is the one from the combination of \textit{Planck pol}, BAO, $H073p0$ and SZ data.

\subsection{The reionization optical depth $\tau$}

The questions of when and how did cosmic reionization take place are still open issues which can be investigated via different cosmological and astrophysical observations. CMB measurements provide the most convincing constraints via the integrated optical depth $\tau$, whose mean value is $\tau \simeq 0.078$, from recent Planck temperature and polarization 2015 measurements~\cite{Ade:2015xua}. In the simplest model of reionization, the so-called \emph{instantaneous} reionization scenario, the former mean values would imply a reionization redshift $8<z_{\textrm{reio}}<10$. However, high-redshift quasar absorption spectra~\cite{Mitra:2015yqa} and observations of Lyman $\alpha$ emitters~\cite{Choudhury:2014uba,Mesinger:2014mqa} seem to conclude that the reionization redshift is $z_{\textrm{reio}}\sim 7$. Due to the fact that these cosmological and astrophysical estimations of the reionization redshift seem to indicate slightly lower values of $\tau$ ($z_{\textrm{reio}}\simeq 7$ would correspond to $\tau=0.05$) than those recently quoted by the Planck collaboration, we shall explore here the impact of a prior based on a lower value of $\tau$. We shall assume in the following priors on $\tau$ of  $0.05\pm 0.01$ and $0.06\pm 0.01$, which would approximately lead to $z_{\textrm{reio}}=7$ and $z_{\textrm{reio}}=8$, and we shall refer to these priors as \emph{tau5} and \emph{tau6} respectively. Tables~\ref{tab:tauprior2} and \ref{tab:tauprior} show the constraints on the total neutrino mass after considering such priors on the reionization optical depth $\tau$ from future cosmological and/or astrophysical measurements. Notice that the most stringent data combination used here (i.e. the one arising from \textit{Planck pol}, BAO, $H073p0$ and SZ measurements) will provide a $95\%$~CL bound on $\sum m_\nu$ of $0.0993 $~eV, assuming a prior on $\tau=0.05\pm 0.01$. Following the latest neutrino oscillation physics analyses~\cite{Gonzalez-Garcia:2014bfa}, the minimum total neutrino mass in the inverted hierarchy is $\sum m_\nu=0.0982 \pm 0.0010$~eV. Since the constraint could be biased by the tension of the H073p0 prior with the Planck results, we repeated the analysis with the same combination of data but excluding this prior. We have found a constraint of $\sum m_\nu<0.0926$~eV at $90\%$~CL., i.e. still hinting for a cosmological tension for the neutrino inverted hierarchy.
Therefore, such a prior on $\tau$ could imply that with current data we are already able to test  the neutrino mass hierarchy, albeit in a not significant way. If future combined measurements of the reionization optical depth agree with the astrophysical expectations, cosmology could offer a window to test the neutrino mass hierarchy.

\section{Conclusions} 
\label{sec:concl}
Cosmological limits on neutrino masses rely strongly on the particular choice of the low redshift observables which are used in combination with Cosmic Microwave Background (CMB) measurements. Here we have examined the different limits in the sum of the three active neutrino masses as well as on the possible extra sterile states (both in its massless and massive versions) arising from different existing priors on the Hubble constant and the cluster mass bias. 

In the Hubble constant case, the prior on $H_0$ from a reanalysis of ~\cite{Efstathiou:2013via}, $H_0=70.6\pm 3.3$ km s$^{-1}$ Mpc$^{-1}$, and dubbed here as $H070p6$, leads to larger upper bounds on $\sum m_\nu$ than the recalibrated value of Riess et al~\cite{Riess:2011yx}  ($H_0=73.0\pm 2.4$ km s$^{-1}$ Mpc$^{-1}$), due to the anticorrelation between the Hubble constant and the neutrino mass. 

When additional sterile neutrino species are also considered in the analyses, the constraints on $\neff$ obtained with the $H070p6$ prior are very similar to those obtained from the combination of CMB measurements and Baryon Acoustic Oscillation data. However, in the case of the other possible $H_0$ prior and neglecting polarisation data, an extra sterile massless or massive neutrino is allowed at the $\sim 2\sigma$ level. Therefore, polarisation measurements are essential to ensure the robustness of the bounds on $\neff$.  

In the case of the Planck Sunyaev-Zeldovich (SZ) Cluster catalog, the crucial prior is the cluster mass bias, taken as constant. If the cluster mass bias parameter is allowed to freely vary, we obtain the tightest $95\%$~CL neutrino mass constraint found here, which is $\sum m_\nu<0.126$~eV, and arises from the combination of \textit{Planck pol}, BAO, the $H073p0$ prior and SZ data.  We have explored as well other estimates of the cluster mass bias, as those coming from weak lensing measurements. For these cases the value of the clustering parameter $\sigma_8$ is smaller and lies $1-2\sigma$ away from the values obtained in the case of a freely varying cluster mass bias. This implies smaller clustering and, consequently,  a larger value of the neutrino mass is allowed: the $95\%$~CL upper bounds on $\sum m_\nu$ range from $0.669$~eV to $\sum m_\nu< 0.531$~eV. Therefore, the tension between the measurements of $\sigma_8$ from Planck and from SZ clusters is translated into a large range of possibilities for $\sum m_\nu$. 

Another tension is that related to CMB and astrophysical measurements of the reionization optical depth. While the former prefers a higher $\tau$ (and consequently, a reionization redshift $z_{\textrm{reio}}\simeq 8-10$), Lyman $\alpha$ and quasar data point to a lower value, $z_{\textrm{reio}}\simeq 7$. Assuming the discrepancy between these two measurements of $\tau$ is solved, we illustrate here the cases $z_{\textrm{reio}}=7$ and $z_{\textrm{reio}}=8$ by imposing priors on $\tau$ of  $0.05\pm 0.01$ and $0.06\pm 0.01$. Interestingly, for the lower prior case, and after combining \textit{Planck pol} with BAO, $H073p0$ and SZ data, a $95\%$ ($90\%$)~CL bound on $\sum m_\nu$ of $0.0993$ ($0.0788$)~eV is obtained, values which lie in the range in which a cosmological measurement of the neutrino mass hierarchy is at reach. This result is robust against the choice of the $H_0$ prior, as using instead the $H_0=70.6\pm 3.3$ km s$^{-1}$ Mpc$^{-1}$ measurement for the Hubble constant with the same data sets quoted above and the \emph{tau5}  prior, the $90\%$~CL  constraint on $\sum m_\nu$ is $0.0962$~eV. Removing the $H_0$ prior would result in a bound of  $\sum m_\nu<0.0926$~eV at $90\%$~CL.

Given the current spread of values in the low redshift priors considered, it is
clearly reasonable to ask if these priors could be of any use for current precision cosmology. In case of the H0 prior,
a better understanding of the several anchors, as discussed
in ~\cite{Efstathiou:2013via}, is mandatory.  
However, considering a large number of priors, without 
preferring one value over another, as we performed in this paper
is probably the best way to present the results and avoid biases from
the choice of a single, low redshift prior.
Moreover, none of the priors considered suggest
the presence of a neutrino mass and they are all compatible with the
results coming from a Planck+BAO analysis that provides the
strongest constraint on neutrino masses.

\vspace{0.2cm}
\section{Acknowledgments} OM is supported by PROMETEO II/2014/050, by the Spanish Grant FPA2011--29678 of the MINECO and by PITN-GA-2011-289442-INVISIBLES. This work has been done within the Labex ILP (reference ANR-10-LABX-63) part of the Idex SUPER, and received financial state aid managed by the Agence Nationale de la Recherche, as part of the programme Investissements d'avenir under the reference ANR-11-IDEX-0004-02. EDV acknowledges the support of the European Research Council via the Grant  number 267117 (DARK, P.I. Joseph Silk).

\begin{table*}
\begin{center}\footnotesize
\scalebox{1.04}{\begin{tabular}{lcccccccc}
\hline \hline
         & Planck & Planck pol & Planck & Planck pol& Planck& Planck pol & Planck &Planck pol   \\                     
         &            &                   & +BAO       &    +BAO &     +H070p6    &+H070p6 &+H073p0 & +H073p0 \\  
\hline
\hspace{1mm}\\

$\Omega_{\textrm{c}}h^2$& $0.1202\,_{-0.0044}^{+0.0047} $& $0.1200\,_{-0.0030}^{+0.0031}$    & $0.1188\,_{-0.0029}^{+0.0028} $& $0.1192\,_{-0.0023}^{+0.0023}$   &$0.1193\,_{-0.0041}^{+0.0042} $& $0.1196\,_{-0.0028}^{+0.0028} $ &$0.1179\,_{-0.0041}^{+0.0040} $ &$0.1189\,_{-0.0028}^{+0.0029} $ \\
\hspace{1mm}\\

$\mnu$ [eV] &      $<0.754$&      $<0.497$ & $<0.220$   & $<0.175$ &$<0.337$& $<0.291$ & $<0.195$ & $<0.180$ \\
\hspace{1mm}\\

$H_0$ &      $65.5\,_{-5.9}^{+4.4}$&      $ 66.3\,_{-3.8}^{+2.9}$ & $ 67.6\,_{-1.3}^{+1.3}$   &  $ 67.5\,_{-1.2}^{+1.1}$&$ 67.1\,_{-3.1}^{+2.8}$& $ 67.0\,_{-2.4}^{+2.1}$ & $ 68.2\,_{-2.3}^{+2.0}$ & $ 67.7\,_{-1.7}^{+1.7}$ \\
\hspace{1mm}\\

$\sigma_8$   & $ 0.79\,_{-0.11}^{+0.08}$   & $ 0.811\,_{-0.076}^{+0.058}$   & $ 0.825\,_{-0.042}^{+0.039}$ &  $ 0.832\,_{-0.034}^{+0.033}$  &$ 0.819\,_{-0.057}^{+0.049}$& $ 0.824\,_{-0.049}^{+0.043}$ & $ 0.829\,_{-0.040}^{+0.038} $&  $ 0.831\,_{-0.036}^{+0.035}$ \\
\hspace{1mm}\\

$\Omega_{\textrm{m}}$ &  $0.340\,_{-0.063}^{+0.088}$ &  $0.329\,_{-0.039}^{+0.052}$  & $0.311\,_{-0.016}^{+0.017}$&  $0.312\,_{-0.014}^{+0.015}$  &$0.318\,_{-0.037}^{+0.041}$& $0.319\,_{-0.027}^{+0.031}$ & $0.304\,_{-0.028}^{+0.029}$ & $0.310\,_{-0.022}^{+0.023}$ \\
\hspace{1mm}\\

$\tau$ & $0.080\,_{-0.038}^{+0.038} $  &  $0.081\,_{-0.034}^{+0.033} $ & $0.082\,_{-0.037}^{+0.038} $ & $0.083\,_{-0.032}^{+0.033} $  &$0.082\,_{-0.037}^{+0.038} $& $0.082\,_{-0.034}^{+0.034} $ & $0.085\,_{-0.038}^{+0.039} $ & $0.083\,_{-0.033}^{+0.032} $ \\
\hspace{1mm}\\

\hline
\hline
\end{tabular}}

\caption{$95\%$~CL constraints on the total neutrino mass and mean values (with their associated $95\%$~CL errors) on other cosmological parameters illustrated here from some of the different combinations of data sets explored in the $\Lambda$CDM+$\mnu$ model, focusing on the effect of the prior on $H_0$.}
\label{tab:mah0}
\end{center}
\end{table*}

\begin{table*}
\begin{center}\footnotesize
\scalebox{1.00}{\begin{tabular}{lccccccc}
\hline \hline
         & Planck & Planck pol & Planck & Planck pol& Planck& Planck pol    \\                     
         &+H068p8 & +H068p8 &   +H070p6ref    &+H070p6ref & + H072p5& +H072p5 \\  
\hline
\hspace{1mm}\\

$\Omega_{\textrm{c}}h^2$& $0.1195\,_{-0.0047}^{+0.0047} $& $0.1198\,_{-0.0028}^{+0.0029}$    & $0.1187\,_{-0.0040}^{+0.0040} $& $0.1194\,_{-0.0028}^{+0.0028}$   &$0.1181\,_{-0.0041}^{+0.0040}$ &$0.1191\,_{-0.0028}^{+0.0028} $\\
\hspace{1mm}\\

$\mnu$ [eV] &      $<0.388$&      $<0.333$ & $<0.265$   & $<0.241$ &$<0.211$& $<0.198$ \\
\hspace{1mm}\\

$H_0$ &      $66.8\,_{-3.4}^{+2.9}$&      $ 66.8\,_{-2.6}^{+2.2}$ & $ 67.5\,_{-2.6}^{+2.4}$   &  $ 67.3\,_{-2.1}^{+1.9}$&$ 68.0\,_{-2.4}^{+2.2}$& $ 67.6\,_{-1.8}^{+1.8}$ \\
\hspace{1mm}\\

$\sigma_8$   & $ 0.815\,_{-0.064}^{+0.053}$   & $ 0.820\,_{-0.054}^{+0.046}$   & $ 0.823\,_{-0.048}^{+0.044}$ &  $ 0.827\,_{-0.043}^{+0.039}$  &$ 0.827\,_{-0.042}^{+0.039}$& $ 0.830\,_{-0.038}^{+0.036}$\\
\hspace{1mm}\\

$\Omega_{\textrm{m}}$ &  $0.322\,_{-0.039}^{+0.045}$ &  $0.321\,_{-0.029}^{+0.034}$  & $0.312\,_{-0.032}^{+0.034}$&  $0.316\,_{-0.025}^{+0.027}$  &$0.306\,_{-0.029}^{+0.030}$& $0.312\,_{-0.023}^{+0.024}$ \\
\hspace{1mm}\\

$\tau$ & $0.080\,_{-0.038}^{+0.039} $  &  $0.081\,_{-0.033}^{+0.033} $ & $0.082\,_{-0.038}^{+0.039} $ & $0.082\,_{-0.033}^{+0.032} $  &$0.084\,_{-0.038}^{+0.039} $& $0.083\,_{-0.033}^{+0.032} $\\
\hspace{1mm}\\

\hline
\hline
\end{tabular}}
\caption{$95\%$~CL constraints on the total neutrino mass and mean values (with their associated $95\%$~CL errors) on other cosmological parameters illustrated here from some of the different combinations of data sets explored in the $\Lambda$CDM+$\mnu$ model, focusing on the effect of the prior on $H_0$.}
\label{tab:mah1}
\end{center}
\end{table*}

\begin{table*}
\begin{center}\footnotesize
\scalebox{0.97}{\begin{tabular}{lcccccccc}
\hline \hline
         & Planck & Planck pol& Planck & Planck pol& Planck   &Planck Pol& Planck   &Planck Pol\\                     
                     &              &     & +BAO       &   +BAO &     +H070p6        &+H070p6 &+H073p0&+H073p0\\  
\hline.
\hspace{1mm}\\

$\Omega_{\textrm{c}}h^2$ &$0.1205\,_{-0.0077}^{+0.0080}$& 
$0.1192\,_{-0.0057}^{+0.0060}$  & 
$0.1212\,_{-0.0071}^{+0.0077}$& 
 $0.1193\,_{-0.0058}^{+0.0062}$  &
 $0.1222\,_{-0.0073}^{+0.0074}$  &
 $0.1190\,_{-0.0060}^{+0.0062}$&
 $0.1235_{-0.0070}^{+0.0071}$&
 $0.1215_{-0.0054}^{+0.0053}$\\
\hspace{1mm}\\

$\mnu$ [eV] &  $<0.796$ &
$<0.582$ &  
$<0.289$ &
 $<0.224$ &
 $<0.417$&
 $<0.365$ &
 $<0.337$&
$<0.249$\\
\hspace{1mm}\\

$\neff $&   $<3.592$
& $<3.359$ &
$<3.636$ &
$<3.384$  &
$<3.707$  &
$<3.374$ &
$<3.961$&
$<3.539$ \\
\hspace{1mm}\\

$H_0$  [km s$^{-1}$ Mpc$^{-1}$] & $64.9_{-8,4}^{+7.2}$&
$65.0_{-5.0}^{+4.4}$ &
$68.4_{-2.8}^{+3.0}$&
$ 67.4_{-2.3}^{+2.4}$ &
$ 68.2_{-4.7}^{+4.6}$ & 
$66.6_{-3.5}^{+3.2}$ &
$70.5_{-4.1}^{+4.2}$ &
$68.2_{-2.8}^{+2.7}$ \\
\hspace{1mm}\\

$\sigma_8$   &$0.781_{-0.119}^{+0.091}$  & 
$ 0.794\,_{-0.085}^{+0.067}$ &  
$0.823\,_{-0.044}^{+0.042}$ &
 $0.823\,_{-0.039}^{+0.037}$ &
 $0.819\,_{-0.062}^{+0.057}$ &
  $ 0.813\,_{-0.055}^{+0.047}$ &
 $ 0.835_{-0.053}^{+0.047}$ &
  $ 0.831_{-0.043}^{+0.038}$\\
\hspace{1mm}\\

$\Omega_{\textrm{m}}$ & $0.351_{-0.080}^{+0.104}$&
 $0.342\,_{-0.046}^{+0.061}$ &
 $0.310\,_{-0.017}^{+0.019}$ &
 $0.315\,_{-0.016}^{+0.017}$ &
  $0.316\,_{-0.043}^{+0.044}$ &
   $0.326\,_{-0.031}^{+0.035}$ &
  $0.298\,_{-0.032}^{+0.034}$ &
 $ 0.313_{-0.024}^{+0.024}$\\
\hspace{1mm}\\

$\tau$ & $0.081_{-0.035}^{+0.035}$& 
  $0.086_{-0.034}^{+0.031}$   & 
  $0.088_{-0.038}^{+0.039}$ &
    $0.079_{-0.046}^{+0.035}$ &
$0.088_{- 0.041}^{+0.044}$ &
  $ 0.083_{-0.035}^{+0.035}$ &
$ 0.098_{-0.041}^{+0.044}$ &
  $ 0.091_{-0.035}^{+0.034}$\\
\hspace{1mm}\\
\hline
\hline
\end{tabular}}
\caption{$95\%$~CL constraints on the total neutrino mass and mean values (with their associated $95\%$~CL errors) on other cosmological parameters illustrated here from some of the different combinations of data sets explored in the $\Lambda$CDM+$\mnu$+$\neff$ model, focusing on the effect of the prior on $H_0$.}
\label{tab:maneffh0}
\end{center}
\end{table*}

\begin{table*}
\begin{center}\footnotesize
\scalebox{0.97}{\begin{tabular}{lcccccccc}
\hline \hline
         & Planck & Planck pol& Planck & Planck pol& Planck   &Planck Pol& Planck   &Planck Pol\\                     
                     &              &     & +BAO       & +  BAO &     +H070p6        &+H070p6 &+H073p0&+H073p0\\  
\hline
\hspace{1mm}\\

$\Omega_{\textrm{c}}h^2$  & $0.1215\,_{-0.0105}^{+0.0090}$ &
$0.1207\,_{-0.0071}^{+0.0061}$ &
$0.1214\,_{-0.0081}^{+0.0081}$ &
$0.1189\,_{-0.0081}^{+0.0068}$   &
 $0.1217\,_{-0.0107}^{+0.0088}$&
 $0.1205\,_{-0.0077}^{+0.0068}$ &
 $0.1235\,_{-0.0082}^{+0.0090}$ &  
  $0.1205\,_{-0.0071}^{+0.0064}$    \\
\hspace{1mm}\\

$\mnu$ [eV] &$<0.676$&  
 $<0.528$    &  
 $<0.263$&
  $<0.199$  &   
  $<0.422$  & 
   $<0.337$&
   $<0.291$&
    $<0.321$  \\
\hspace{1mm}\\

$\ms$ [eV] & $<0.972$  &
  $<0.820$  &
  $<0.449$&
 $<0.694$  &
 $<0.822$  &
 $<0.773$&
 $<0.462$&
 $<0.630$
 \\
\hspace{1mm}\\

$\neff $&   $<3.648$  &
 $<3.401$ &
 $<3.762$&
 $<3.405$ & 
 $<3.705$& 
 $<3.445$&
 $<3.961$& 
 $<3.434$  \\
\hspace{1mm}\\

$H_0$[km s$^{-1}$ Mpc$^{-1}$]  & $65.7_{-6.1}^{+5.7}$ &
$65.5_{-3.7}^{+3.2}$ &  
$67.7_{-1.6}^{+1.8}$  & 
$68.7_{-2.4}^{+2.8}$  & 
$67.4_{-4.2}^{+4.4}$ &
$66.5_{-2.8}^{+2.7}$&
$70.0_{-4.2}^{+4.6}$ &
$67.4_{-2.1}^{+2.3}$ 
\\
\hspace{1mm}\\

$\sigma_8$  & $ 0.762\,_{-0.107}^{+0.095}$ &
$ 0.768\,_{-0.087}^{+0.077}$ &
$ 0.801\,_{-0.058}^{+0.051}$ &
$0.806\,_{-0.054}^{+0.048}$  & 
$0.786_{-0.083}^{+0.076}$ &
 $ 0.785\,_{-0.075}^{+0.066}$&
 $ 0.818\,_{-0.068}^{+0.064}$&
   $ 0.803\,_{-0.062}^{+0.056}$\\
\hspace{1mm}\\

$\Omega_{\textrm{m}}$ &  $ 0.350\,_{-0.069}^{+0.083}$ &
$0.347\,_{-0.045}^{+0.054}$ & 
$0.311_{-0.017}^{+0.017}$& 
$0.316\,_{-0.015}^{+0.015}$ & 
$0.328_{-0.045}^{+0.051}$  &
$0.334\,_{-0.034}^{+0.037}$&
$0.305_{-0.037}^{+0.038}$  &
$0.323_{-0.027}^{+0.023}$  
\\
\hspace{1mm}\\

$\tau$ & $0.088_{-0.041}^{+0.043}$ & 
$0.087_{-0.036}^{+0.035}$ &
 $0.095_{-0.040}^{+0.041}$ &
$0.089_{-0.034}^{+0.034}$  & 
 $0.090_{-0.040}^{+0.042}$  &
 $0.087_{-0.035}^{+0.035}$&
 $0.103_{-0.044}^{+0.043}$  &
 $0.091_{-0.035}^{+0.036}$  
   \\
\hspace{1mm}\\
\hline
\hline
\end{tabular}}
\caption{$95\%$~CL constraints on the total neutrino mass and mean values (with their associated $95\%$~CL errors) on other cosmological parameters illustrated here from some of the different combinations of data sets explored in the $\Lambda$CDM+$\mnu$+$\neff$+$\ms$ model, focusing on the effect of the prior on $H_0$.}
\label{tab:maneffmsh0}
\end{center}
\end{table*}

\begin{table*}
\begin{center}\footnotesize
\scalebox{0.93}{\begin{tabular}{lccccccccc}
\hline \hline
                         & Planck  & Planck pol& Planck & Planck pol& Planck & Planck pol& Planck+BAO & Planck pol+BAO \\   
                         &   +SZ    &  +SZ         &  +BAO+SZ &    +BAO+SZ  &        +H073p0+SZ   &+H073p0+SZ & +H073p0+SZ   &+H073p0+SZ & \\  
\hline
\hspace{1mm}\\

$\Omega_{\textrm{c}}h^2$ & $0.1182\,_{-0.0041}^{+0.0041}$& $0.1191\,_{-0.0028}^{+0.0028}$  &$0.1185\,_{-0.0027}^{+0.0026}$  & $0.1189\,_{-0.0022}^{+0.0022}$ &  $0.1168\,_{-0.0038}^{+0.0039}$ & $0.1183\,_{-0.0027}^{+0.0027}$
& $0.1180\,_{-0.0027}^{+0.0025}$  &  $0.1186\,_{-0.0021}^{+0.0021}$   \\
\hspace{1mm}\\

$\mnu$ [eV] &      $<0.206$&      $<0.184$&      $<0.175$ & $<0.147$  &$<0.139$ &$<0.129$& $<0.155$ &$<0.126$ \\
\hspace{1mm}\\

$H_0$  [km s$^{-1}$ Mpc$^{-1}$]  &      $68.0\,_{-2.4}^{+2.2}$ &      $67.6\,_{-1.8}^{+1.9}$ &$67.9\,_{-1.2}^{+1.2}$&  $67.8\,_{-1.1}^{+1.0}$   &$68.8\,_{-1.9}^{+1.9}$ &$68.2\,_{-1.5}^{+1.3}$
& $68.2\,_{-1.2}^{+1.1}$ &$68.0\,_{-1.0}^{+1.0}$\\
\hspace{1mm}\\

$\sigma_8$   & $ 0.830\,_{-0.041}^{+0.039}$  & $ 0.834\,_{-0.036}^{+0.034}$ & $ 0.831\,_{-0.039}^{+0.034}$  &  $ 0.835\,_{-0.032}^{+0.029}$ &$0.834\,_{-0.036}^{+0.033}$  &$0.837\,_{-0.031}^{+0.030}$ 
 &   $ 0.832\,_{-0.037}^{+0.036}$ &$0.837\,_{-0.031}^{+0.028}$ \\
\hspace{1mm}\\

$\Omega_{\textrm{m}}$ &  $0.306\,_{-0.029}^{+0.030}$&  $0.311\,_{-0.022}^{+0.023}$&  $0.307\,_{-0.015}^{+0.016}$  & $0.309\,_{-0.013}^{+0.014}$  & $0.295\,_{-0.023}^{+0.025}$ & $0.304\,_{-0.018}^{+0.019}$&  $0.303\,_{-0.014}^{+0.015}$ & $0.306\,_{-0.012}^{+0.013}$ \\
\hspace{1mm}\\

$\tau$ & $0.087\,_{-0.037}^{+0.038} $& $0.085\,_{-0.034}^{+0.033} $ &$0.085\,_{-0.036}^{+0.036} $ &  $0.085\,_{-0.033}^{+0.032} $  &$0.092\,_{-0.037}^{+0.038} $ &$0.088\,_{-0.034}^{+0.032} $ & $0.086\,_{-0.036}^{+0.036} $ &$0.087\,_{-0.033}^{+0.032} $ \\
\hspace{1mm}\\
\hline
\hline
\end{tabular}}

\caption{As Table~\ref{tab:mah0}, but including measurements from the Planck SZ Cluster Catalog.}
\label{tab:mapsz}
\end{center}
\end{table*}

\begin{table*}
\begin{center}\footnotesize
\scalebox{1.00}{\begin{tabular}{lccccccc}
\hline \hline
         & Planck & Planck pol & Planck & Planck pol& Planck& Planck pol    \\                     
         &+SZ+H068p8 & +SZ+H068p8 &   +SZ+H070p6ref    &+SZ+H070p6ref & +SZ+ H072p5& +SZ+H072p5 \\  
\hline
\hspace{1mm}\\

$\Omega_{\textrm{c}}h^2$& $0.1181\,_{-0.0039}^{+0.0040} $& $0.1190\,_{-0.0027}^{+0.0028}$    & $0.1176\,_{-0.0038}^{+0.0039} $& $0.1187\,_{-0.0027}^{+0.0027}$   &$0.1170\,_{-0.0038}^{+0.0039}$ &$0.1184\,_{-0.0027}^{+0.0027} $\\
\hspace{1mm}\\

$\mnu$ [eV] &      $<0.189$&      $<0.179$ & $<0.164$   & $<0.154$ &$<0.145$& $<0.136$ \\
\hspace{1mm}\\

$H_0$ &      $66.1\,_{-2.2}^{+2.0}$&      $ 67.7\,_{-1.7}^{+1.6}$ & $ 68.4\,_{-2.0}^{+1.9}$   &  $ 67.9\,_{-1.6}^{+1.5}$&$ 68.7\,_{-2.0}^{+1.9}$& $ 68.1\,_{-1.5}^{+1.5}$ \\
\hspace{1mm}\\

$\sigma_8$   & $ 0.831\,_{-0.039}^{+0.037}$   & $ 0.834\,_{-0.035}^{+0.034}$   & $ 0.832\,_{-0.037}^{+0.036}$ &  $ 0.836\,_{-0.033}^{+0.032}$  &$ 0.833\,_{-0.037}^{+0.033}$& $ 0.837\,_{-0.031}^{+0.031}$\\
\hspace{1mm}\\

$\Omega_{\textrm{m}}$ &  $0.305\,_{-0.027}^{+0.028}$ &  $0.310\,_{-0.021}^{+0.022}$  & $0.301\,_{-0.025}^{+0.026}$&  $0.307\,_{-0.019}^{+0.020}$  &$0.297\,_{-0.023}^{+0.026}$& $0.305\,_{-0.019}^{+0.020}$ \\
\hspace{1mm}\\

$\tau$ & $0.087\,_{-0.037}^{+0.038} $  &  $0.085\,_{-0.034}^{+0.033} $ & $0.089\,_{-0.037}^{+0.038} $ & $0.086\,_{-0.034}^{+0.033} $  &$0.091\,_{-0.037}^{+0.038} $& $0.087\,_{-0.034}^{+0.033} $\\
\hspace{1mm}\\

\hline
\hline
\end{tabular}}
\caption{As Table~\ref{tab:mah1}, but including measurements from the Planck SZ Cluster Catalog.}
\label{tab:mapsz2}
\end{center}
\end{table*}

\begin{table*}
\begin{center}\footnotesize
\scalebox{1.00}{\begin{tabular}{lcccccc}
\hline \hline
          &  Planck & Planck pol& Planck   & Planck pol& Planck& Planck pol \\                     
                     & +SZ (CCCP)  & +SZ (CCCP)   & +SZ (WtG) & +SZ (WtG) & +SZ (CMBlens) & +SZ (CMBlens)  \\  
\hline
\hspace{1mm}\\

$\Omega_{\textrm{c}}h^2$ & $0.1190_{-0.0041}^{+0.0041}$ &
 $0.1197_{-0.0028}^{+0.0029}$ & 
 $0.1193_{-0.0038}^{+0.0037}$&   
 $0.1198_{-0.0028}^{+0.0028}$  &
$0.1182_{-0.0039}^{+0.0040}$ &
  $ 0.1193_{-0.0027}^{+0.0027}$ \\
\hspace{1mm}\\

$\mnu$ [eV] &   $<0.542$&
 $<0.576$ &
  $<0.506$&  
  $<0.531$ &
 $<0.634$ &
  $<0.669$\\
\hspace{1mm}\\

$H_0$  [km s$^{-1}$ Mpc$^{-1}$] &  $65.8_{-3.2}^{+3.1}$&
$65.5_{-2.9}^{+2.7}$ &
 $65.9_{-3.0}^{+2.8} $ &
 $65.6_{-2.8}^{+2.5}$ & 
$65.7_{-3.4}^{+3.3}$ &
 $65.1_{-3.0}^{+2.8}$ \\
\hspace{1mm}\\

$\sigma_8$   &$0.780_{-0.061}^{+0.056}$ &
$ 0.783_{-0.064}^{+0.056}$  & 
$0.788_{-0.057}^{+0.050}$&   
$0.789_{-0.057}^{+0.048}$ &   
$0.763_{-0.061}^{+0.059}$&
$ 0.764_{-0.064}^{+0.062}$ \\
\hspace{1mm}\\

$\Omega_{\textrm{m}}$ &$0.333_{-0.042}^{+0.045}$ &
$0.338_{-0.036}^{+0.042}$  & 
$0.333_{-0.038}^{+0.042}$ &
 $ 0.337_{0.035}^{+0.040}$   &
$0.334_{-0.045}^{+0.049}$ &
 $ 0.343_{-0.041}^{+0.043}$  \\
\hspace{1mm}\\

$\tau$ &$0.077_{-0.037}^{+0.037}$ &
$0.079_{-0.034}^{+0.034}$     & 
$0.079_{-0.037}^{+0.038}$  & 
$0.080_{-0.033}^{+0.033}$ & 
$0.074_{-0.039}^{+0.039}$&
$0.075_{-0.035}^{+0.036}$ \\
\hspace{1mm}\\

\hline
\hline
\end{tabular}}
\caption{$95\%$~CL constraints on the total neutrino mass and mean values (with their associated $95\%$~CL errors) on other cosmological parameters illustrated here from some of the different combinations of data sets explored in the $\Lambda$CDM+$\mnu$ model, focusing on the effect of the prior on the cluster mass bias, see text for details.}
\label{tab:macluster}
\end{center}
\end{table*}

\begin{table*}
\begin{center}\footnotesize
\scalebox{0.95}{\begin{tabular}{lcccccccc}
\hline \hline
                          & Planckl& Planck pol&      Planck           & Planck pol   &Planck                        & Planck pol   &Planck+BAO                        & Planck pol+BAO  \\   
                           &+SZ       &  +SZ      &     +BAO+SZ   &  +BAO+SZ      &+SZ+H073p0           &+SZ+H073p0                  & +H073p0+SZ  &+H073p0+SZ \\  
\hline
\hspace{1mm}\\

$\Omega_{\textrm{c}}h^2$ &$0.1209\,_{-0.0078}^{+0.0082}$&
$0.1192\,_{-0.0057}^{+0.0059}$&
$0.1209_{-0.0077}^{+0.0077}$&
$0.1195_{-0.0061}^{+0.0062}$&
$0.1237_{-0.0072}^{+0.0073}$&
$0.1214\,_{-0.0054}^{+0.0055}$&
$0.1235_{-0.0070}^{+0.0070}$&
$0.1217\,_{-0.0058}^{+0.0058}$ \\
\hspace{1mm}\\

$\mnu$ [eV]  &$<0.434$&
$<0.374$&
$<0.299$&
$<0.205$&
$<0.326$&
$<0.253$&
$<0.331$&  
 $<0.200$\\
\hspace{1mm}\\

$\neff$ & $<3.635$ &
$<3.313$ &
$<3.648$&
$<3.389$&
$<3.858$ &
$3.571$&
$<3.800$&
 $<3.561$\\
\hspace{1mm}\\
 
$H_0$ [km s$^{-1}$ Mpc$^{-1}$] & $66.8_{-4.9}^{+5.3}$    &
$65.8_{-3.5}^{+3.4}$    &
$68.3_{-3.0}^{+3.0}$&
$67.4_{-2.3}^{+2.4}$&
$70.3_{-3.6}^{+3.8}$&
$68.2\,_{-2.7}^{+3.0}$&    
 $69.7_{-2.5}^{+2.6}$&
$68.5\,_{-2.1}^{+2.0}$\\
\hspace{1mm}\\

$\sigma_8$  &$ 0.807_{-0.063}^{+0.060}$ &
$ 0.806_{-0.054}^{+0.049}$ &
$0.822_{-0.046}^{+0.042}$&
$0.824_{0.037}^{+0.036}$&
$0.833_{-0.046}^{+0.048}$&
$0.312\,_{-0.024}^{+0.024}$&
$0.831_{-0.046}^{+0.042}$&
  $ 0.834\,_{-0.034}^{+0.035}$ \\
\hspace{1mm}\\

$\Omega_{\textrm{m}}$ & $0.327_{-0.047}^{+0.049}$&
 $0.332_{-0.033}^{+0.036}$&
$0.310_{-0.017}^{+0.018}$ &
$0.315_{-0.015}^{+0.016}$ &
$0.300_{-0.029}^{+0.030}$&
$0.312\,_{-0.024}^{+0.024}$&
$0.304_{-0.016}^{+0.016}$& 
$0.309\,_{-0.013}^{+0.014}$ \\
\hspace{1mm}\\

$\tau$ & $0.084_{-0.040}^{+0.041}$&
$0.081_{-0.035}^{+0.035}$&
$0.089_{-0.037}^{+0.038}$&
$0.085_{-0.034}^{+0.033}$&
$0.095_{-0.040}^{+0.040}$&
$0.087\,_{-0.032}^{+0.032}$&
$0.094_{-0.037}^{+0.037}$&
$0.089\,_{-0.033}^{+0.034}$ \\
\hspace{1mm}\\

\hline
\hline
\end{tabular}}
\caption{As Table~\ref{tab:maneffh0}, but including measurements from the Planck SZ Cluster Catalog.}
\label{tab:maneffpsz}
\end{center}
\end{table*}

\begin{table*}
\begin{center}\footnotesize
\scalebox{0.95}{\begin{tabular}{lcccccccc}
\hline \hline
                          & Planckl& Planck pol&      Planck           & Planck pol   &Planck                        & Planck pol &Planck+BAO                       & Planck pol+BAO  \\   
                           &+SZ       &  +SZ      &     +BAO+SZ   &  +BAO+SZ      &SZ+H073p0   &SZ+H073p0              & +H073p0+SZ  &+H073p0+SZ \\  
\hline
\hspace{1mm}\\

$\Omega_{\textrm{c}}h^2$ &$0.1220\,_{-0.0080}^{+0.0085}$ &
 $0.1207\,_{-0.0060}^{+0.0057}$ &
 $0.1215_{-0.0075}^{+0.0075}$&
 $0.1191\,_{-0.0078}^{+0.0070}$& 
 $0.1235_{-0.0077}^{+0.0079}$&
$0.1210_{-0.0063}^{+0.0065}$ &
 $0.1237_{-0.0078}^{+0.0081}$ & 
 $ 0.1204_{-0.0075}^{+0.0070}$  \\
\hspace{1mm}\\

$\mnu$ [eV] & $<0.370$&
$<0.362$  &
$<0.265$&
$<0.191$&   
 $<0.297$&
$<0.217$ &
$<0.275$ & 
$<0.190$ \\
\hspace{1mm}\\

$\ms$ [eV] &  $<0.640$&
$<0.630$  &    
 $<0.356$&
$<0.659$   &
$<0.385$&
$<0.512$&
$<0.330$&      
$<0.506$\\
\hspace{1mm}\\

$\neff$ & $<3.666$&
$<3.412$ &
$<3.723$&
$<3.405$&  
$<3.860$&
$<3.525$&
  $<3.894$ &  
$<3.478$  \\
\hspace{1mm}\\

$H_0$ [km s$^{-1}$ Mpc$^{-1}$]& $67.1_{-3.6}^{+4.0}$  & 
$66.3_{-2.8}^{+2.4}$  &  
$68.7_{-2.4}^{+2.5}$&
$67.7_{-1.5}^{+1.8}$   &
$69.7_{-3.5}^{+3.5}$&
$67.7_{-2.3}^{+2.6}$&
$69.7_{-1.5}^{+1.6}$&  
$68.3_{-1.7}^{+1.8}$\\
\hspace{1mm}\\

$\sigma_8$  & $ 0.789_{-0.066}^{+0.061}$  &
$ 0.786_{-0.065}^{+0.059}$  &
$0.808_{-0.051}^{+0.050}$&
$ 0.808\,_{-0.052}^{+0.047}$& 
 $0.815_{-0.057}^{+0.055}$&   
 $0.808_{-0.057}^{+0.053}$&
$0.816_{-0.051}^{+0.046}$ &
$0.817_{-0.051}^{+0.047}$  \\
\hspace{1mm}\\

$\Omega_{\textrm{m}}$ & $0.331_{-0.041}^{+0.043}$ &
$0.336_{-0.032}^{+0.037}$ &
$0.332_{-0.017}^{+0.017}$&
$0.316\,_{-0.014}^{+0.015}$ &  
$0.308_{-0.031}^{+0.034}$& 
$0.320_{-0.026}^{+0.026}$&
$0.307_{-0.015}^{+0.016}$&
 $0.312\,_{-0.013}^{+0.015}$ \\
\hspace{1mm}\\

$\tau$ &  $ 0.090_{-0.039}^{+0.041}$ &
$ 0.088_{-0.035}^{+0.040}$ &
$0.094_{-0.038}^{+0.040}$&
$0.090\,_{-0.036}^{+0.034} $& 
 $0.100_{-0.042}^{+0.040}$ &
  $0.092_{-0.034}^{+0.033}$&
 $0.098_{-0.039}^{+0.040}$ &
$0.091\,_{-0.034}^{+0.033} $ \\
\hspace{1mm}\\

\hline
\hline
\end{tabular}}
\caption{As Table~\ref{tab:maneffmsh0}, but including measurements from the Planck SZ Cluster Catalog.}
\label{tab:maneffmspsz}
\end{center}
\end{table*}

\begin{table*}
\footnotesize
\scalebox{0.98}
{\begin{tabular}{lccccccc}
\hline \hline
                         &  Planck pol& Planck pol& Planck pol+BAO& Planck pol+BAO& Planck pol+BAO& Planck pol  +BAO\\   
                         &  +BAO+tau6 &  +BAO+tau5  &    +H072p5+tau6  &    +H072p5+tau5 &  +H072p5+SZ+tau6 &  +H072p5+SZ+tau5  \\  
                         
\hline
\hspace{1mm}\\

$\Omega_{\textrm{c}}h^2$ & $0.1196\,_{-0.0026}^{+0.0027}$ & $0.1198\,_{-0.0020}^{+0.0023}$ & $0.1194\,_{-0.0021}^{+0.0021}$ &$0.1195\,_{-0.0020}^{+0.0022}$ & $0.1191\,_{-0.0020}^{+0.0020}$ & $0.1192\,_{-0.0021}^{+0.0020}$ \\
\hspace{1mm}\\

$\mnu$ [eV] &      $<0.141$ &      $<0.128$ &   $<0.122$   & $<0.116$&      $<0.107$ &      $<0.101$\\
\hspace{1mm}\\

$H_0$  [km s$^{-1}$ Mpc$^{-1}$] &      $67.4\,_{-1.1}^{+1.0}$ &      $67.4\,_{-1.0}^{+1.0}$ &   $67.6\,_{-1.0}^{+1.0}$   &$67.6\,_{-1.0}^{+1.0}$ &      $67.84\,_{-0.99}^{+0.96}$ &      $67.79\,_{-0.98}^{+0.96}$ \\
\hspace{1mm}\\

$\sigma_8$   & $ 0.822\,_{-0.027}^{+0.024}$ & $ 0.818\,_{-0.025}^{+0.023}$ &$ 0.823\,_{-0.025}^{+0.022}$ &$ 0.818\,_{-0.024}^{+0.022}$ & $ 0.824\,_{-0.022}^{+0.021}$ & $ 0.819\,_{-0.022}^{+0.021}$  \\
\hspace{1mm}\\

$\Omega_{\textrm{m}}$ &  $0.313\,_{-0.013}^{+0.014}$ &  $0.314\,_{-0.013}^{+0.014}$ &$0.311\,_{-0.013}^{+0.014}$  & $0.311\,_{-0.012}^{+0.014}$ &  $0.308\,_{-0.012}^{+0.013}$ &  $0.309\,_{-0.013}^{+0.013}$ \\
\hspace{1mm}\\

$\tau$ & $0.066\,_{-0.017}^{+0.017} $ & $0.059\,_{-0.017}^{+0.017} $ & $0.066\,_{-0.017}^{+0.017} $& $0.059\,_{-0.018}^{+0.017} $ & $0.067\,_{-0.017}^{+0.017} $ & $0.059\,_{-0.017}^{+0.017} $\\
\hspace{1mm}\\
\hline
\hline
\end{tabular}}
\caption{$95\%$~CL constraints on the total neutrino mass and mean values (with their associated $95\%$~CL errors)  on other cosmological parameters illustrated here from some of the different combinations of data sets explored in the $\Lambda$CDM+$\mnu$ model, focusing on the effect of the prior on the reionization optical depth $\tau$, see text for details.}
\label{tab:tauprior2}
\end{table*}

\begin{table*}
\footnotesize
\scalebox{0.98}
{\begin{tabular}{lccccccc}
\hline \hline
                         &  Planck pol& Planck pol& Planck pol& Planck pol& Planck pol+BAO& Planck pol  +BAO\\   
                         &  +BAO+SZ+tau6 &  +BAO+SZ+tau5  &    H073p0+SZ+tau6  &    H073p0+SZ+tau5 &  +H073p0+SZ+tau6 &  +H073p0+SZ+tau5  \\  
                         
\hline
\hspace{1mm}\\

$\Omega_{\textrm{c}}h^2$ & $0.1194\,_{-0.0021}^{+0.0021}$ & $0.1195\,_{-0.0021}^{+0.0021}$ & $0.1190\,_{-0.0025}^{+0.0026}$ &$0.1192\,_{-0.0025}^{+0.0026}$ & $0.1190\,_{-0.0020}^{+0.0020}$ & $0.1192\,_{-0.0021}^{+0.0020}$ \\
\hspace{1mm}\\

$\mnu$ [eV] &      $<0.122$ &      $<0.116$ &   $<0.112$   & $<0.107$&      $<0.104$ &      $<0.0993$\\
\hspace{1mm}\\

$H_0$  [km s$^{-1}$ Mpc$^{-1}$] &      $67.7\,_{-1.0}^{+1.0}$ &      $67.6\,_{-1.0}^{+1.0}$ &   $67.9\,_{-1.4}^{+1.3}$   &$67.8\,_{-1.4}^{+1.2}$ &      $67.88\,_{-0.98}^{+0.96}$ &      $67.83\,_{-0.98}^{+0.99}$ \\
\hspace{1mm}\\

$\sigma_8$   & $ 0.823\,_{-0.024}^{+0.022}$ & $ 0.818\,_{-0.023}^{+0.022}$ &$ 0.824\,_{-0.023}^{+0.022}$ &$ 0.819\,_{-0.022}^{+0.021}$ & $ 0.824\,_{-0.022}^{+0.021}$ & $ 0.819\,_{-0.022}^{+0.021}$  \\
\hspace{1mm}\\

$\Omega_{\textrm{m}}$ &  $0.311\,_{-0.013}^{+0.013}$ &  $0.311\,_{-0.013}^{+0.014}$ &$0.307\,_{-0.017}^{+0.018}$  & $0.309\,_{-0.017}^{+0.018}$ &  $0.308\,_{-0.012}^{+0.013}$ &  $0.308\,_{-0.013}^{+0.013}$ \\
\hspace{1mm}\\

$\tau$ & $0.066\,_{-0.017}^{+0.017} $ & $0.059\,_{-0.017}^{+0.017} $ & $0.067\,_{-0.017}^{+0.017} $& $0.060\,_{-0.017}^{+0.017} $ & $0.067\,_{-0.017}^{+0.017} $ & $0.059\,_{-0.017}^{+0.017} $\\
\hspace{1mm}\\
\hline
\hline
\end{tabular}}
\caption{$95\%$~CL constraints on the total neutrino mass and mean values (with their associated $95\%$~CL errors)  on other cosmological parameters illustrated here from some of the different combinations of data sets explored in the $\Lambda$CDM+$\mnu$ model, focusing on the effect of the prior on the reionization optical depth $\tau$, see text for details.}
\label{tab:tauprior}
\end{table*}




\begin{thebibliography}{99}

\bibitem{Ade:2015xua} 
  P.~A.~R.~Ade {\it et al.} [Planck Collaboration],
  arXiv:1502.01589 [astro-ph.CO].
\bibitem{DiValentino:2015wba}
  E.~Di Valentino, E.~Giusarma, M.~Lattanzi, O.~Mena, A.~Melchiorri and J.~Silk,
  arXiv:1507.08665 [astro-ph.CO].
\bibitem{Giusarma:2014zza} 
  E.~Giusarma, E.~Di Valentino, M.~Lattanzi, A.~Melchiorri and O.~Mena,
  Phys.\ Rev.\ D {\bf 90}, 043507 (2014)
  [arXiv:1403.4852 [astro-ph.CO]].

\bibitem{Gerbino:2015ixa}
  M.~Gerbino, M.~Lattanzi and A.~Melchiorri,
  arXiv:1507.08614 [hep-ph].

\bibitem{Palanque-Delabrouille:2015pga} 
  N.~Palanque-Delabrouille, C.~Yeche, J.~Baur, C.~Magneville, G.~Rossi, J.~Lesgourgues, A.~Borde and E.~Burtin {\it et al.},
  arXiv:1506.05976 [astro-ph.CO].

\bibitem{Slosar:2006xb} 
  A.~Slosar,
  Phys.\ Rev.\ D {\bf 73}, 123501 (2006)
  doi:10.1103/PhysRevD.73.123501
  [astro-ph/0602133].

\bibitem{Adam:2015rua} 
  R.~Adam {\it et al.}  [Planck Collaboration],
  arXiv:1502.01582 [astro-ph.CO].

\bibitem{Beutler:2011hx} 
  F.~Beutler, C.~Blake, M.~Colless, D.~H.~Jones, L.~Staveley-Smith, L.~Campbell, Q.~Parker and W.~Saunders {\it et al.},
  Mon.\ Not.\ Roy.\ Astron.\ Soc.\  {\bf 416}, 3017 (2011)
  [arXiv:1106.3366 [astro-ph.CO]].


\bibitem{Ross:2014qpa} 
  A.~J.~Ross, L.~Samushia, C.~Howlett, W.~J.~Percival, A.~Burden and M.~Manera,
  Mon.\ Not.\ Roy.\ Astron.\ Soc.\  {\bf 449}, no. 1, 835 (2015)
  [arXiv:1409.3242 [astro-ph.CO]].

\bibitem{Anderson:2013zyy} 
  L.~Anderson {\it et al.}  [BOSS Collaboration],
  Mon.\ Not.\ Roy.\ Astron.\ Soc.\  {\bf 441}, no. 1, 24 (2014)
  [arXiv:1312.4877 [astro-ph.CO]].

\bibitem{Lewis:2002ah}
  A.~Lewis and S.~Bridle,
  Phys.\ Rev.\ D {\bf 66}, 103511 (2002)
  [astro-ph/0205436].

\bibitem{Lewis:2013hha} 
  A.~Lewis,
  Phys.\ Rev.\ D {\bf 87}, no. 10, 103529 (2013)
  [arXiv:1304.4473 [astro-ph.CO]].
  
\bibitem{Aghanim:2015wva} 
  N.~Aghanim {\it et al.} [Planck Collaboration],
  ``Planck 2015 results. XI. CMB power spectra, likelihoods, and robustness of parameters,''
  \eprint{arXiv:1507.02704}.



  \bibitem{Humphreys:2013eja} 
  E.~M.~L.~Humphreys, M.~J.~Reid, J.~M.~Moran, L.~J.~Greenhill and A.~L.~Argon,
  Astrophys.\ J.\  {\bf 775}, 13 (2013)
  [arXiv:1307.6031 [astro-ph.CO]].

\bibitem{Riess:2011yx} 
  A.~G.~Riess {\it et al.},
  Astrophys.\ J.\  {\bf 730}, 119 (2011)
  [Astrophys.\ J.\  {\bf 732}, 129 (2011)]
  [arXiv:1103.2976 [astro-ph.CO]].
  \bibitem{Bennett:2014tka}
  C.~L.~Bennett, D.~Larson, J.~L.~Weiland and G.~Hinshaw,
  Astrophys.\ J.\  {\bf 794} (2014) 135
  [arXiv:1406.1718 [astro-ph.CO]].
  \bibitem{Cuesta:2014asa} 
  A.~J.~Cuesta, L.~Verde, A.~Riess and R.~Jimenez,
  Mon.\ Not.\ Roy.\ Astron.\ Soc.\  {\bf 448}, no. 4, 3463 (2015)
  [arXiv:1411.1094 [astro-ph.CO]].
\bibitem{Efstathiou:2013via} 
  G.~Efstathiou,
  Mon.\ Not.\ Roy.\ Astron.\ Soc.\  {\bf 440}, no. 2, 1138 (2014)
  [arXiv:1311.3461 [astro-ph.CO]].

\bibitem{Rigault:2014kaa} 
  M.~Rigault {\it et al.},
  Astrophys.\ J.\  {\bf 802}, no. 1, 20 (2015)
  doi:10.1088/0004-637X/802/1/20
  [arXiv:1412.6501 [astro-ph.CO]].
    

  
   \bibitem{Giusarma:2012ph} 
  E.~Giusarma, R.~De Putter and O.~Mena,
  Phys.\ Rev.\ D {\bf 87}, no. 4, 043515 (2013)
  [arXiv:1211.2154 [astro-ph.CO]].
  \bibitem{Abazajian:2012ys} 
  K.~N.~Abazajian {\it et al.},
  arXiv:1204.5379 [hep-ph].
  \bibitem{Melchiorri:2008gq} 
  A.~Melchiorri, O.~Mena, S.~Palomares-Ruiz, S.~Pascoli, A.~Slosar and M.~Sorel,
  JCAP {\bf 0901}, 036 (2009)
  [arXiv:0810.5133 [hep-ph]].
   
 \bibitem{Ade:2015gva} 
   P.~A.~R.~Ade {\it et al.} [Planck Collaboration],
   arXiv:1502.01598 [astro-ph.CO].
 \bibitem{Ade:2015fva} 
   P.~A.~R.~Ade {\it et al.} [Planck Collaboration],
   arXiv:1502.01597 [astro-ph.CO].
  
  
   \bibitem{Tinker:2008ff} 
  J.~L.~Tinker, A.~V.~Kravtsov, A.~Klypin, K.~Abazajian, M.~S.~Warren, G.~Yepes, S.~Gottlober and D.~E.~Holz,
  Astrophys.\ J.\  {\bf 688}, 709 (2008)
  [arXiv:0803.2706 [astro-ph]].
 
\bibitem{vonderLinden:2014haa} 
  A.~von der Linden {\it et al.},
  Mon.\ Not.\ Roy.\ Astron.\ Soc.\  {\bf 443}, no. 3, 1973 (2014)
  [arXiv:1402.2670 [astro-ph.CO]].

 \bibitem{Zaldarriaga:1998te} 
  M.~Zaldarriaga and U.~Seljak,
  Phys.\ Rev.\ D {\bf 59}, 123507 (1999)
  [astro-ph/9810257].
  \bibitem{Melin:2014uaa} 
  J.~B.~Melin and J.~G.~Bartlett,
  Astron.\ Astrophys.\  {\bf 578}, A21 (2015)
  [arXiv:1408.5633 [astro-ph.CO]].
  \bibitem{Mitra:2015yqa} 
  S.~Mitra, T.~R.~Choudhury and A.~Ferrara,
  arXiv:1505.05507 [astro-ph.CO].
\bibitem{Choudhury:2014uba} 
  T.~R.~Choudhury, E.~Puchwein, M.~G.~Haehnelt and J.~S.~Bolton,
  arXiv:1412.4790 [astro-ph.CO].
  \bibitem{Mesinger:2014mqa} 
  A.~Mesinger, A.~Aykutalp, E.~Vanzella, L.~Pentericci, A.~Ferrara and M.~Dijkstra,
  Mon.\ Not.\ Roy.\ Astron.\ Soc.\  {\bf 446}, 566 (2015)
  [arXiv:1406.6373 [astro-ph.CO]].
\bibitem{Gonzalez-Garcia:2014bfa} 
  M.~C.~Gonzalez-Garcia, M.~Maltoni and T.~Schwetz,
  JHEP {\bf 1411}, 052 (2014)
  [arXiv:1409.5439 [hep-ph]].
  
  
\end{thebibliography}
\end{document}